# Estimating entanglement in a class of $N$-qudit states

Sumiyoshi Abe[1,2,3]

[1]*Physics Division, College of Information Science and Engineering,*
*Huaqiao University, Xiamen 361021, China*
[2]*Department of Physical Engineering, Mie University, Mie 514-8507, Japan*
[3]*Institute of Physics, Kazan Federal University, Kazan 420008, Russia*

**Abstract.** The logarithmic derivative (or, quantum score) of a positive definite density matrix appearing in the quantum Fisher information is discussed, and its exact expression is presented. Then, the problem of estimating the parameters in a class of the Werner-type $N$-qudit states is studied in the context of the quantum Cramér-Rao inequality. The largest value of the lower bound to the error of estimate by the quantum Fisher information is shown to coincide with the separability point only in the case of two qubits. It is found, on the other hand, that such largest values give rise to the universal fidelity that is independent of the system size.

Quantum states are characterized by a set of parameters, in general, and accordingly it is of significant importance to estimate their values as precisely as possible, leading to the quantum-mechanical counterpart of statistical estimation theory [1-4]. Recent examples are estimations of outputs of quantum channels [5], coupling parameters in quantum critical systems [6], parameters in quantum optical states [7], and thermal degree of freedom in thermofield dynamics [8], to name a few. Today, its fundamental importance for quantum information technology is widely appreciated [9,10].

Let $\rho(\theta)$ be a density matrix of the system under consideration and dependent on a parameter, $\theta$. Here, we require $\rho(\theta)$ to be positive definite, although a generic density matrix can be positive semidefinite. A task is to estimate the values of the parameter through measuring a relevant quantity of the system. Let such a quantity be an observable, $A$. It has a spectral decomposition, $A = \sum_\alpha \alpha E_\alpha$, where the collection, $\{E_\alpha\}_\alpha$, forms a positive operator-valued measure, that is, $E_\alpha \geq 0$ and $\sum_\alpha E_\alpha = I$ with $I$ being the identity operator. Outcome, $\alpha$, occurs with the probability $p_\alpha(\theta) = \mathrm{tr}[E_\alpha \rho(\theta)]$, which satisfies the normalization condition, $\sum_\alpha p_\alpha(\theta) = 1$. Let $\hat{\theta}(A)$ be an unbiased estimator:

$$\langle \hat{\theta}(A) \rangle = \theta, \tag{1}$$

which should hold, at $\theta$, for the expectation value of the observable, $\hat{\theta}(A) = \sum_\alpha \hat{\theta}(\alpha) E_\alpha$, where $\langle \hat{\theta}(A) \rangle = \mathrm{tr}[\hat{\theta}(A) \rho(\theta)] = \sum_\alpha \hat{\theta}(\alpha) p_\alpha(\theta)$. The mean square error of estimation is given by the variance

$$\left(\delta\hat{\theta}\right)^2 = \sum_\alpha p_\alpha(\theta)\left[\hat{\theta}(\alpha)-\theta\right]^2. \qquad (2)$$

Then, the Cramér-Rao inequality states that

$$\left(\delta\hat{\theta}\right)^2 \geq \frac{1}{J_A[p(\theta)]}, \qquad (3)$$

where $J_A[p(\theta)]$ is the Fisher information defined by

$$J_A[p(\theta)] = \sum_\alpha p_\alpha(\theta)\left[\frac{\partial \ln p_\alpha(\theta)}{\partial \theta}\right]^2. \qquad (4)$$

Given a problem, it is generally hard to identify an observable $A$ to be measured (or, equivalently, an associated positive operator-valued measure) that can minimize the right-hand side of Eq. (3). Accordingly, the quantum Fisher information,

$$J[\rho(\theta)] = \text{tr}\left[\rho(\theta) L^2(\theta)\right], \qquad (5)$$

as the upper bound of $J_A[p(\theta)]$ is often examined. This quantity gives an approximate scale of errors. $L$, in Eq. (5) stands for the logarithmic derivative (or, quantum score) of the density matrix, $\rho(\theta)$. Widely employed are the symmetric and right logarithmic derivatives defined by

$$\frac{\partial \rho(\theta)}{\partial \theta} = \frac{1}{2}\left[\rho(\theta) L^{(S)}(\theta) + L^{(S)}(\theta) \rho(\theta)\right], \tag{6}$$

$$\frac{\partial \rho(\theta)}{\partial \theta} = \rho(\theta) L^{(R)}(\theta), \tag{7}$$

respectively. Clearly, the right logarithmic derivative is not Hermitian, in general. Accordingly, the right-hand side of Eq. (5) need be modified as $\text{tr}\left\{\left[L^{(R)}(\theta)\right]^\dagger \rho(\theta) L^{(R)}(\theta)\right\}$, for example.

Our purpose here is to examine the scheme for estimating entanglement. To our knowledge, estimation of entanglement has recently been discussed in Refs. [11,12]. The authors of Ref. [11] have used the estimation scheme for an interpretation of entanglement as a resource for parameter estimation. The discussion in Ref. [12] has been made about estimating the amount of entanglement. In contrast to these works, here we examine the scheme in connection with the separability condition on a class of mixed states of multipartite systems. In particular, we consider a class of $N$-qudit states, i.e., the Werner-type states of an $N$ partite $d$-level system and estimate the values of the parameters describing strength of correlations. We evaluate the quantum Fisher information in Eq. (5) and analyze its property. We find a remarkable result that the fidelity, $F$, between the Werner-type states and the maximally entangled states at the values of the parameter yielding the minimum quantum Fisher information is

$$F = \frac{1}{2}, \tag{8}$$

which is universal in the sense that it does not depend on the system size.

Before discussing the problem of estimating entanglement, here we make a brief comment on the logarithmic derivative of a positive definite density matrix. The symmetric and right logarithmic derivatives in Eqs. (6) and (7) are clearly useful because of their simplicities. However, they are not the rigorous logarithmic derivative, in general. To directly calculate the logarithmic derivative, it is useful to employ the following representation:

$$\ln \rho(\theta) = \int_0^\infty \frac{dx}{x} \left( e^{-x} - e^{-x\rho(\theta)} \right). \tag{9}$$

Thus, the logarithmic derivative is transformed to the familiar derivative of an exponential operator:

$$\frac{\partial e^{A(\theta)}}{\partial \theta} = \int_0^1 d\lambda \, e^{\lambda A(\theta)} \frac{\partial A(\theta)}{\partial \theta} e^{(1-\lambda)A(\theta)} \tag{10}$$

with $A(\theta) = -x\rho(\theta)$, in which the formula $e^X Y e^{-X} = Y + [X,Y] + (1/2!)[X,[X,Y]] + \cdots$ can be used. This gives the exact expression for the quantum score. The present manipulation will afford the quantity in Eq. (9) a compact expression, if $\rho(\theta)$ possesses a Lie-algebraic structure, for example.

Now, let us address ourselves to the problem of estimating entanglement in the $N$-qudit Werner-type states.

Firstly, we consider the case of the smallest system that is a bipartite 2-level system. Its Werner state [13,14] reads

$$\rho(\theta) = \frac{1-\theta}{4} I_1^{(2)} \otimes I_2^{(2)} + \theta |\Psi\rangle\langle\Psi|, \tag{11}$$

where $\theta \in [0,1)$, $|\Psi\rangle = (|0\rangle_1 |0\rangle_2 + |1\rangle_1 |1\rangle_2)/\sqrt{2}$, and $I_i^{(2)}$ ($i=1,2$) denotes the $2\times 2$ identity operator in the space of the $i$-th particle. This is a convex combination of the completely random state and the maximally entangled state and can experimentally be generated [15-17]. It is known [18,19] (see also Ref. [20]) that the separation point between separable and entangled states is at $\theta^* = 1/3$, that is, the state in Eq. (11) is separable if $\theta < 1/3$. (Recall that the total density matrix, which is expressed as a convex combination of the tensor products of the marginal density matrices of all particles, is called separable, and correlations contained in such a density matrix is not of entanglement and can be described by models with local hidden variables.) The derivative and inverse of the density matrix are given by $\partial \rho(\theta)/\partial \theta = (-1/4) I_1^{(2)} \otimes I_2^{(2)} + |\Psi\rangle\langle\Psi|$ and $\rho^{-1}(\theta) = [4/(1-\theta)] I_1^{(2)} \otimes I_2^{(2)} - 16\theta/[(1-\theta)(1+3\theta)] |\Psi\rangle\langle\Psi|$, respectively. Since these commute with each other, the expressions for the logarithmic derivative of the density matrix in Eqs. (6), (7), and (10) coincide with each other. $J[\rho(\theta)]$ in Eq. (5) is then calculated to be $J[\rho(\theta)] = 3/[(1-\theta)(1+3\theta)]$. The divergence in the limit $\theta \to 1-$ essentially comes from the absence of the inverse of a pure-state density matrix. The minimum is at $\theta = 1/3$, which coincides with the separation point, $\theta^*$. Therefore, a lower bound on the error of estimation evaluated by $1/J[\rho(\theta)]$ becomes largest at the separation point.

It turns out, however, that the coincidence mentioned above occurs only in the case of the a bipartite 2-level system. To see this, let us consider the $N$-qudit Werner-type state, which reads

$$\rho(\theta) = \frac{1-\theta}{d^N} I_1^{(d)} \otimes I_2^{(d)} \otimes \cdots \otimes I_N^{(d)} + \theta |\Phi\rangle\langle\Phi|, \qquad (12)$$

where $\theta \in [0,1)$, $I_i^{(d)}$ ($i = 1, 2, ..., N$) is the $d \times d$ identity operator, and $|\Phi\rangle = (1/\sqrt{d}) \sum_{m=0}^{d-1} |m\rangle_1 |m\rangle_2 \cdots |m\rangle_N$ with $\{|m\rangle_i\}_{m=0,1,...,d-1}$ being a complete orthonormal system in the space of the $i$-th particle with $d$ levels. The separation point of this state is known to be at $\theta^* = 1/(1+d^{N-1})$ [21-24]. Again, the derivative and inverse of the density matrix, $\partial \rho(\theta)/\partial \theta = (-1/d^N) I_1^{(d)} \otimes I_2^{(d)} \otimes \cdots \otimes I_N^{(d)} + |\Phi\rangle\langle\Phi|$ and $\rho^{-1}(\theta) = [d^N/(1-\theta)] I_1^{(d)} \otimes I_2^{(d)} \otimes \cdots \otimes I_N^{(d)} - d^{2N}\theta/\{(1-\theta)[1+(d^N-1)\theta]\} |\Phi\rangle\langle\Phi|$, are seen to commute with each other. The quantum Fisher information in Eq. (5) is calculated to be

$$J[\rho(\theta)] = \frac{d^N - 1}{(1-\theta)[1+(d^N-1)\theta]}. \qquad (13)$$

Its minimum is seen to be at

$$\theta = \frac{d^N - 2}{2(d^N - 1)}, \qquad (14)$$

which is equal to or larger than the separation point, $\theta^*$ ($\theta = \theta^*$ iff $N = 2$ and $d = 2$, i.e., the bipartite 2-level system). In other words, the values of $\theta$ in Eq. (14) are always in the ranges, in which the states are entangled.

The result in Eq. (14) has an implication of interest. $1/J[\rho(\theta)]$ with the value of $\theta$ in Eq. (14) tells us the scale of errors in estimation. Since entanglement is not associated with local observables, a quantity to be considered for the state in Eq. (12) may be the fidelity [25,26] (see also Refs. [27,28]), which in the present case is given by $F = F[\sigma, \rho(\theta)] = \left[ \text{tr} \left( \sqrt{\sigma} \, \rho(\theta) \sqrt{\sigma} \right)^{1/2} \right]^2$ with the maximally entangled state, $\sigma = |\Phi\rangle\langle\Phi|$. Its value is found to be

$$F = \frac{1 - \theta + d^N \theta}{d^N}. \tag{15}$$

The allowed range of $F$ in consistency with that of $\theta$ is $[1/d^N, 1)$. Substituting the value of $\theta$ in Eq. (14) into this expression, we obtain the result in Eq. (8).

In conclusion, we have discussed the problem of entanglement in a class of $N$-qudit states in the context of parameter estimation. We have found that the value of the parameter at which the quantum Fisher information takes its minimum is always in the range where the state is entangled and remarkably the corresponding value of the fidelity is independent of the system size.

**Acknowledgements**

This work has been supported in part by grants from National Natural Science Foundation of China (No. 11775084) and Grant-in-Aid for Scientific Research from the Japan Society for the Promotion of Science (No. 26400391 and No. 16K05484), and by

**References**

[1] Helstrom C.W. Quantum Detection and Estimation Theory. New York, Academic Press, 1976.

[2] Holevo A.S. Probabilistic and Statistical Aspects of Quantum Theory. Amsterdam, North-Holland, 1982.

[3] Barndorff-Nielsen O.E., Gill R.D., Jupp P.E. On quantum statistical inference. *J. R. Statist. Soc. B*, 2003, vol. 65, no. 4, pp. 775-816. doi: 10.1111/1467-9868.00415.

[4] Petz D., Ghinea C. Introduction to quantum Fisher information. *Quantum Probability and Related Topics*, R. Rebolledo, M. Orszag (eds.). Singapore, World Sccientific, 2011, pp. 261-281. doi: 10.1142/9789814338745_0015.

[5] Sarovar M., Milburn G.J. Optimal estimation of one-parameter quantum channels. *J. Phys. A: Math. Gen.*, 2006, vol. 39, no. 26, pp. 8487-8505. doi: 10.1088/0305-4470/39/26/015.

[6] Zanardi P., Paris M.G.A., Venuti L.C. Quantum criticality as a resource for quantum estimation. *Phys. Rev. A*, 2008, vol. 78, no. 4, 042105–1-7. doi: 10.1103/PhysRevA.78.042105.

[7] Bradshaw M., Assad S.M., Lam P.K. A tight Cramér-Rao bound for joint parameter estimation with a pure two-mode squeezed probe. *Phys. Lett. A*, 2017,

vol. 381, no. 32, pp. 2598-2607. doi: 10.1016/j.physleta.2017.06.024.

[8] Abe S. Estimation of the thermal degree of freedom in thermo field dynamics. *Phys. Lett. A*, 1999, vol. 254, no. 3-4, pp. 149-153. doi: 10.1016/S0375-9601(99)00061-4.

[9] *Asymptotic Theory of Quantum Statistical Inference*, M. Hayashi (ed.). Singapore, World Scientific, 2005.

[10] Wiseman H.M., Milburn G.J. Quantum Measurement and Control. Cambridge, Cambridge Univ. Press, 2010.

[11] Boixo S., Monras A. Operational interpretation for global multipartite entanglement. *Phys. Rev. Lett.*, 2008, vol. 100, no. 10, 100503–1-4. doi: 10.1103/PhysRevLett.100.100503.

[12] Genoni M.G., Giorda P., Paris M.G.A. Optimal estimation of entanglement. *Phys. Rev. A*, 2008, vol. 78, no. 3, 032303–1-9. doi: 10.1103/PhysRevA.78.032303.

[13] Werner R.F. Quantum states with Einstein-Podolsky-Rosen correlations admitting a hidden-variable model. *Phys. Rev. A*, 1989, vol. 40, no. 8, pp. 4277-4281. doi: 10.1103/PhysRevA.40.4277.

[14] Popescu S. Bell's inequalities versus teleportation: What is nonlocality? *Phys. Rev. Lett.*, 1994, vol. 72, no. 6, pp. 797-799. doi: 10.1103/PhysRevLett.72.797.

[15] Zhang Y.-S., Huang Y.-F., Li C.-F., Guo G.-C. Experimental preparation of the Werner state via spontaneous parametric down-conversion. *Phys. Rev. A*, 2002,

vol. 66, no. 6, 062315–1-4. (2002). doi: 10.1103/PhysRevA.66.062315.

[16] Barbieri M., De Martini F., Di Nepi G., Mataloni P., D'Ariano G.M., Macchiavello C. Detection of entanglement with polarized photons: Experimental realization of an entanglement witness. *Phys. Rev. Lett.*, 2003, vol. 91, no. 22, 227901–1-4. doi: 10.1103/PhysRevLett.91.227901.

[17] Jakóbczyk L. Generation of Werner-like stationary states of two qubits in a thermal reservoir. *J. Phys. B: At. Mol. Opt. Phys.*, 2010, vol. 43, no. 1, 015502–1-7. doi: 10.1088/0953-4075/43/1/015502.

[18] Peres A. Separability criterion for density matrices. *Phys. Rev. Lett.*, 1996, vol. 77, no. 8, pp. 1413-1415. doi: 10.1103/PhysRevLett.77.1413.

[19] Horodecki M., Horodecki P., Horodecki R. Separability of mixed states: Necessary and sufficient conditions. *Phys. Lett. A*, 1996, vol. 223, no. 1-2, pp. 1-8. doi: 10.1016/S0375-9601(96)00706-2.

[20] Abe S., Rajagopal A.K. Nonadditive conditional entropy and its significance for local realism. *Physica A*, 2001, vol. 289, no. 1-2, pp. 157-164. doi: 10.1016/S0378-4371(00)00476-3.

[21] Pittenger A.O., Rubin M.H. Separability and Fourier representations of density matrices. *Phys. Rev. A*, 2000, vol. 62, no. 3, 032313–1-9. doi: 10.1103/PhysRevA.62.032313.

[22] Pittenger A.O., Rubin M.H. Note on separability of the Werner states in arbitrary dimensions. *Opt. Commun.*, 2000, vol. 179, no. 1-6, pp. 447-449. doi: 10.1016/S0030-4018(00)00612-X.

[23] Abe S. Nonadditive information measure and quantum entanglement in a class of mixed states of an $N^n$ system. *Phys. Rev. A*, 2002, vol. 65, no. 5, 052323–1-6. doi: 10.1103/PhysRevA.65.052323.

[24] Nayak A.S., Sudha, Usha Devi A.R., Rajagopal A.K. One parameter family of *N*-qudit Werner-Popescu states: Bipartite separability using conditional quantum relative Tsallis entropy. *J. Quantum Inf. Sci.*, 2018, vol. 8, no. 1, pp. 12-23. doi: 10.4236/jqis.2018.81002.

[25] Jozsa R. Fidelity for mixed quantum states. *J. Mod. Opt.*, 1994, vol. 41, no. 12, pp. 2315-2323. doi: 10.1080/09500349414552171.

[26] Schumacher B. Quantum coding. *Phys. Rev. A*, 1995, vol. 51, no. 4, pp. 2738-2747. doi: 10.1103/PhysRevA.51.2738.

[27] Abe S. Nonadditive generalization of the quantum Kullback-Leibler divergence for measuring the degree of purification. *Phys. Rev. A*, 2003, vol. 68, no. 3, 032302–1-3. doi: 10.1103/PhysRevA.68.032302.

[28] Rajagopal A.K., Usha Devi A.R., Rendell R.W. Kraus representation of quantum evolution and fidelity as manifestations of Markovian and non-Markovian forms. *Phys. Rev. A*, 2010, vol. 82, no. 4, 042107–1-7. doi: 10.1103/PhysRevA.82.042107.